\documentclass[aps,twocolumn,showpacs]{revtex4}
\usepackage{graphicx}
\usepackage{epsfig}
\usepackage{epstopdf}
\usepackage{amsfonts}
\usepackage{amssymb}
\usepackage{amsbsy}
\usepackage{amsmath}
\usepackage{mathrsfs}
\usepackage{latexsym}
\usepackage{natbib}
\usepackage{bm}
\usepackage{color}
\usepackage{braket}
\usepackage{slashed}
\usepackage{pgfplots}

\DeclareMathOperator{\arcsinh}{asinh}

\usepackage{tikz}
\usetikzlibrary{shapes,arrows,shadows}

\def\x{\mathbf{x}}

\def\k{\mathrm{k}}

\def\muimb{\bar{\mu}_{Im}}
\def\Ith{I^{\text{th}}}

\def\sigmab{\bar{\sigma}}
\def\omegab{\bar{\omega}}

\def\gbsigma{\bar{g}_{\sigma}}
\def\gtsigma{\tilde{g}_{\sigma}}

\def\gbomega{\bar{g}_{\omega}}
\def\gtomega{\tilde{g}_{\omega}}

\def\mr{\mathcal{M}}

\begin{document}

\author{Golam Mortuza Hossain}
\email{ghossain@iiserkol.ac.in}

\author{Susobhan Mandal}
\email{sm17rs045@iiserkol.ac.in}

\affiliation{ Department of Physical Sciences, 
Indian Institute of Science Education and Research Kolkata,
Mohanpur - 741 246, WB, India }
 
\pacs{04.62.+v, 26.60.+c, 97.60.Jd}

\date{\today}

\title{Higher mass limits of neutron stars from the equation 
of states in curved spacetime}

\begin{abstract}
In order to solve the Tolman-Oppenheimer-Volkoff equations for neutron stars, 
one routinely uses the equation of states which are computed in the Minkowski 
spacetime. Using a first-principle approach, it is shown that the equation of 
states which are computed within the curved spacetime of the neutron stars 
include the effect of gravitational time dilation. It arises due to the 
radially varying interior metric over the length scale of the star and 
consequently it leads to a much higher mass limit. As an example, for a given 
set of parameters in a $\sigma-\omega$ model of nuclear matter, the maximum 
mass limit is shown to increase from $1.61 M_{\odot}$ to $2.24 M_{\odot}$ due to 
the inclusion of gravitational time dilation.
\end{abstract}

\maketitle

\section{Introduction}

A key lesson of Einstein's general relativity is that a curved spacetime 
can be described locally by the Minkowski metric \emph{i.e.} a curved spacetime 
is locally flat. This argument is then often used to deploy the equation of 
states which are computed in the Minkowski spacetime, for solving the 
Tolman-Oppenheimer-Volkoff (TOV) equations in the study of neutron stars. 
Here we refer such an equation of state (EOS) as a \emph{flat} EOS. It may be 
noted that the TOV eqs. follow from Einstein's equation for a spherically 
symmetric interior geometry in the general relativity.

However, such an approach overlooks the fact that two locally inertial frames 
which are located at different radial positions within the star, are \emph{not} 
identical as these frames are subject to the \emph{gravitational time dilation}. 
In other words, while the spacetime metric of these two frames are locally flat, 
the corresponding clock rates are not the same as they have different 
lapse functions. This aspect directly impacts the respective matter field 
dynamics. After all, as famously stated by Wheeler, in general relativity not 
only matter tells spacetime how to curve but also spacetime tells matter how to 
move.

Recently, by employing a first-principle approach, we gave a derivation of the 
equation of state for an ensemble of non-interacting degenerate neutrons in the 
interior curved spacetime of a spherical star \cite{hossain2021equation}. We 
refer such an equation of state as a \emph{curved} EOS. For regular stars the 
effect of gravitational time dilation on their matter field dynamics is 
negligible. However, for the compact stars, such as the neutron stars, the 
effect of gravitational time dilation is significant. In particular, we have 
shown in the ref. \cite{hossain2021equation} that for a neutron star containing 
non-interacting ideal degenerate neutrons, the maximum mass limit increases from 
$0.71 M_{\odot}$ to $0.83 M_{\odot}$. Clearly, the usage of flat EOS leads one 
to underestimate the maximum mass limit.

Nevertheless, the consideration of non-interacting degenerate neutron matter 
alone is not sufficient to describe the matter contents of the 
\emph{astrophysical} neutron stars whose maximum mass limits have now been 
observed to be more than $2 M_{\odot}$ \cite{shapiro2008black, 
nattila2017neutron, ozel2016dense}. In order to explain the observed mass-radius 
relation of the neutron stars, various kinds of nuclear matter EOS have been 
studied in the literature \cite{shen2002complete, lattimer2016equation, 
douchin2001unified, tolos2016equation, maieron2004hybrid, klahn2007modern, 
baldo2007quark}. These models are inspired by different particle physics models 
and generally include different types of interactions between the possible 
nucleons for describing the nuclear matter contained within the astrophysical 
neutron stars. Among them, a widely used model is known as the so-called 
$\sigma-\omega$ model \cite{whittenbury2014quark, katayama2012equation, 
miyatsu2013new} that includes interacting baryons, leptons, mesons, and often a 
set of hyperons \cite{chatterjee2016hyperons, schaffner2008hypernuclear, 
balberg1999roles, dhapo2010appearance, bednarek2001structure, 
hornick2018relativistic}.

In order to perform a first-principle derivation of the equation of states using 
interior curved spacetime of the neutron stars, here we consider a simplified 
$\sigma-\omega$ model containing the neutron, the proton and the electron as the 
fermions and a scalar meson $\sigma$ and a self-interacting vector meson 
$\omega$. Subsequently, we show that the EOS which is derived in the curved 
spacetime, incorporates the effect of gravitational time dilation. The EOS 
reduces exactly to its flat spacetime counterpart when the effect of time 
dilation is turned-off. By considering different sets of parameters, we 
show that the maximum mass limits of the neutron stars always increase due to 
the inclusion of general relativistic time dilation.

This article is organized as follows. In the section 
\ref{sec:InteriorSpacetime}, we briefly review the form of the spherically 
symmetric interior metric of the stars. In the section 
\ref{sec:SigmaOmegaModel}, we study the $\sigma-\omega$ model in detail. In 
particular, by computing the grand canonical partition function, we derive the 
equation of state corresponding to the $\sigma-\omega$ model by using the 
interior curved spacetime. In the section \ref{sec:NumericalEvaluation}, by 
employing numerical methods, we study the properties of the curved EOS and then 
compare it with the corresponding flat EOS. Subsequently, we solve the TOV eqs. 
numerically to obtain the mass-radius relations of the neutron stars. In the 
section \ref{sec:UniversalityTimeDilation}, we discuss about the general 
transformations that are required to obtain the curved EOS starting from the 
corresponding flat EOS. We conclude the article with the discussions in the 
section \ref{sec:Discussions}.

\section{Interior metric of spherical stars}\label{sec:InteriorSpacetime}

The invariant distance element inside a spherically symmetric star can be 
expressed as
\begin{equation}\label{InteriorMetric}
ds^2 = - e^{2\Phi(r)}dt^2 + e^{2\nu(r)} dr^2 + 
r^2(d\theta^2 + \sin^2\theta d\phi^2) ~,
\end{equation}
where we have used the so-called \emph{natural units} \emph{i.e.} the speed of 
light $c$ and Plank's constant $\hbar$ are set to unity. In general relativity, 
the metric functions $\Phi(r)$ and $\nu(r)$ are governed by Einstein's equation 
and the conservation equation of the stress-energy tensor. These equations, 
known as the TOV equations, are given by 
\begin{equation}\label{TOVEqn}
\frac{d\Phi}{dr} = \frac{G(\mr + 4\pi r^3 P)}{r(r - 2 G \mr)} ~~,~~
\frac{dP}{dr} = - (\rho + P) \frac{d\Phi}{dr} ~,
\end{equation}
where $d\mr = 4\pi r^2\rho dr$, $P$ is the pressure and $\rho$ is the energy 
density. Additionally, the equation for the metric function $\nu(r)$ can be 
partially solved to obtain a relation $e^{-2\nu(r)} = (1 - 2 G \mr/r)$. 

\section{$\sigma-\omega$ model of nuclear matter}\label{sec:SigmaOmegaModel}

In the framework of quantum hadrodynamics (QHD) \cite{serot1992quantum, 
serot1997recent}, the $\sigma-\omega$ model is a well-known model which is often 
used to describe the nuclear matter contained within the neutron stars. In order 
to study the effect of gravitational time dilation on the properties of the 
equation of states, here we consider a simplified $\sigma -\omega$ model 
containing two \emph{baryons}, namely the neutron and the proton, a 
\emph{lepton} namely the electron, a massive \emph{scalar} meson $\sigma$ and a 
self-interacting \emph{vector} meson $\omega$. The corresponding action in an 
arbitrary curved spacetime with a metric $g_{\mu\nu}$ can be expressed as 
\begin{equation}\label{SigmaOmegaTotalAction}
S = \int d^{4}x \sqrt{-g} ~ \mathcal{L} = \int d^{4}x \sqrt{-g} 
\left[ \mathcal{L}_{D} + \mathcal{L}_{M} \right]  ~,
\end{equation}
where $g$ is the metric determinant. The Lagrangian density for the 
Dirac fermions is
\begin{equation}\label{FermionLagrangianDensity}
\mathcal{L}_{D} = - \sum_{I=n,p,e}\bar{\psi}_{I}
(i {e^{\mu}}_{a} \gamma^{a}\mathcal{D}_{\mu}+m_{I})\psi_{I} ~,
\end{equation}
where ${e^{\mu}}_{a}$ are the \emph{tetrad} components and $\mathcal{D}_{\mu}$ 
is the covariant derivative of the spinor fields \cite{hossain2021equation}. 
The indices $n$, $p$ and $e$ refer to the neutron, the proton and the electron 
respectively. Here Dirac matrices $\gamma^{a}$ satisfy 
$\{\gamma^{a},\gamma^{b}\} = - 2\eta^{ab} \mathbb{I}$ with $\eta^{ab}$ being 
the Minkowski metric.
The total Lagrangian density for the mesons $\sigma$ and $\omega$ can be 
expressed as $\mathcal{L}_{M} = \mathcal{L}_{\sigma} + \mathcal{L}_{\sigma i} 
+ \mathcal{L}_{\omega} + \mathcal{L}_{\omega i}$, where the Lagrangian density 
for the free $\sigma$ meson is
\begin{equation}\label{SigmaActionFree}
 \mathcal{L}_{\sigma}  = -\frac{1}{2} 
g^{\mu\nu}\partial_{\mu}\sigma\partial_{\nu}\sigma 
-\frac{1}{2} m_{\sigma}^{2}\sigma^{2} ~,
\end{equation}
and its interaction with the baryons is described by
\begin{equation}\label{SigmaActionInteraction}
\mathcal{L}_{\sigma i} = \sum_{I=n,p} 
g_{\sigma}\sigma \, \bar{\psi}_{I}\psi_{I} ~,
\end{equation}
with $g_{\sigma}$ being the dimensionless coupling constant. Similarly, the 
Lagrangian density for the free vector meson $\omega$ is given by
\begin{equation}\label{OmegaActionFree}
 \mathcal{L}_{\omega} = - g^{\mu\rho} g^{\nu\lambda} 
 (\nabla_{[\mu}\omega_{\nu]}) (\nabla_{[\rho}\omega_{\lambda]})
 - \frac{1}{2} m_{\omega}^{2}g^{\mu\nu}\omega_{\mu}\omega_{\nu} ~,
\end{equation}
where $\nabla_{[\mu}\omega_{\nu]} = \tfrac{1}{2} (\nabla_{\mu}\omega_{\nu} - 
\nabla_{\nu}\omega_{\mu})$. The self-interaction of $\omega$ meson and its 
interaction with the baryons are described by
\begin{equation}\label{OmegaActionInteraction}
\mathcal{L}_{\omega i} = \frac{\zeta~g_{\omega}^{4}}{4!} 
(g^{\mu\nu}\omega_{\mu}\omega_{\nu})^{2} +
\sum_{I=n,p} g_{\omega}\omega_{\mu} {e^{\mu}}_{a} 
\bar{\psi}_{I}\gamma^{a}\psi_{I}  ~,
\end{equation}
where $g_{\omega}$ is the coupling constant with the baryons and $\zeta$ is the 
coupling constant controlling its quartic self-interaction. Both of these 
coupling constants here are dimensionless.

\subsection{Reduced action for fermions}

Inside a star, the pressure $P$ and the energy density $\rho$ both vary 
radially. On the other hand, at a thermodynamical equilibrium, these quantities 
are required to be uniform within the system of interest. In order to reconcile 
these two apparently conflicting aspects, one needs to consider a 
sufficiently small spatial region around every given point such that the 
physical quantities within the small region can be considered to be spatially 
uniform. At the same, the small region must also contain sufficiently large 
number of degrees of freedom for achieving the thermal equilibrium. In other 
words, around every point inside the star the notion of a \emph{local} 
thermodynamical equilibrium must hold.

We have mentioned earlier that one can always find a locally flat coordinate 
system around any given point. Here we follow the ref. 
\cite{hossain2021equation}, to construct such a locally flat coordinate system 
which also retains the information about the corresponding lapse function. Let 
us consider a small box around a point which is located at a radial location 
$r_0$. In order to achieve a local thermodynamical equilibrium, the metric 
functions $\Phi(r)$ and $\nu(r)$ within the box can be approximated to be 
uniform. For a spherical star we can always rotate the polar axis such that it 
passes through the center of the box. It ensures $\theta$ to be small for every 
points within the box. Subsequently, by defining a new set of coordinates
$X = e^{\nu(r_0)} r \sin \bar{\theta} \cos\phi$, 
$Y = e^{\nu(r_0)} r \sin \bar{\theta} \sin\phi$, and
$Z = e^{\nu(r_0)} r \cos \bar{\theta}$ along with $\bar{\theta} =  
e^{-\nu(r_0)}\theta$, one can reduce the metric inside the box to be
\begin{equation}\label{MetricInTOVBox}
ds^2 = - e^{2\Phi(r_0)}dt^2 + dX^2 + dY^2 + dZ^2 ~.
\end{equation}
We note that the metric (\ref{MetricInTOVBox}) despite being flat within the 
box, is not \emph{globally} flat as the radial location of the box in the
metric function $\Phi(r_0)$ is retained. In other words, for the purpose of 
computing the equation of states of the nuclear matter within the box, 
$\Phi(r_0)$ has a fixed value whereas in the TOV eqs. (\ref{TOVEqn}) the 
same metric function can be seen to be varying radially, as like the 
pressure $P(r_0)$ and the energy density $\rho(r_0)$ within the box.

By assuming a diagonal form, the tetrad and the inverse tetrad corresponding to 
the metric (\ref{MetricInTOVBox}), can be written as
\begin{equation}\label{TetradInTOVBox}
{e_{\mu}}^{a} = diag(e^{\Phi},1,1,1)  ~,~
{e^{\mu}}_{a} = diag(e^{-\Phi},1,1,1)  ~,
\end{equation}
where $\Phi \equiv \Phi(r_0)$. As a consequence, the associated 
spin-connections of the spin-covariant derivative $\mathcal{D}_{\mu}$ or 
equivalently Ricci rotation coefficients \cite{dvornikov2019neutrino} of the
tetrad fields vanish and the spin-covariant derivative becomes 
$\mathcal{D}_{\mu} = \partial_{\mu}$. The diagonal ansatz for the tetrad 
(\ref{TetradInTOVBox}) fixes the gauge freedoms that arise due to the usage 
tetrad field. In general, the tetrad field in 4 dimension has 16 
components at each spacetime point in contrast to the metric field 
which has only 10 components. The Dirac action describing the free 
fermions, then reduces to
\begin{equation}\label{FermionActionReduced}
S_D = -\int d^{4}x \sum_I \bar{\psi}_I \left[ i \gamma^{0} \partial_0 
+ e^{\Phi} \left(i \gamma^{k} \partial_{k} + m_I\right) \right]\psi_I  ~,
\end{equation}
where the index $k$ runs over $1,2,3$. The conserved charge, corresponding to 
the 4-current $j^{\mu}_I$ of the $\Ith$ fermion, then becomes $Q_I = \int 
d^3x \sqrt{-g} j^0_I = \int d^3x \bar{\psi}_I \gamma^{0} \psi_I$.

The reduced action (\ref{FermionActionReduced}) within the box can be viewed as 
a modified spinor action written in the Minkowski spacetime with metric 
$diag(-1,1,1,1)$. It contains the information about the global metric 
function $\Phi$, unlike the spinor action which is written in a globally flat 
spacetime and routinely used in the literature. We employ this modified 
spinor action for subsequent analysis.

\subsection{RMF approximation for mesons}\label{sec:rmf}

In QHD, the coupling strengths of interactions are not necessarily weak. 
In other words, a potential perturbation series will diverge unless an 
appropriate resummation method \cite{andersen2005resummation, sterman2003multi} 
is applied but application of such a method is often not feasible. Further, any 
approximation method that one employs is required to be accurate for both lower 
and higher baryon number densities. In this context, such aims are usually 
achieved by using the so-called \emph{relativistic mean field} (RMF) 
approximation \cite{ring1996relativistic, mueller1996relativistic}.

In the RMF approximation, one replaces the meson field operators by their vacuum 
expectation values which are then treated as the classical fields. On the other 
hand, the vacuum expectation values of the kinetic terms and the spatial 
components $\langle\hat{\omega}^{i}\rangle \equiv  \bra{0} \hat{\omega}^{i} 
\ket{0}$ vanish as these expectation values within the box should be both 
uniform and stationary to ensure local thermodynamical equilibrium. In summary, 
for the meson fields, the RMF approximation leads to
\begin{equation}\label{RMFApproximation}
 \hat{\omega}^{\mu} \rightarrow \langle\hat{\omega}^{\mu}\rangle
 = \langle\hat{\omega}^0\rangle \delta_{0}^{\mu} 
 = \langle \hat{\omega}_0\rangle g^{0\mu}  
 ~~,~~ \hat{\sigma} \rightarrow \langle\hat{\sigma} \rangle ~.
\end{equation}
In principle, the coupling constants $g_{\sigma}$ and $g_{\omega}$ can have 
different values for different baryons. However, for simplicity here we
assume that these  coupling constants have same values for both neutrons 
and protons. The Euler-Lagrange equation for the $\sigma$ meson then becomes
\begin{equation}\label{EoMSigma}
 \sigmab \equiv  m_{\sigma} \langle\hat{\sigma}\rangle 
 =  \gbsigma \sum_{I=n,p} n_I^S  ~,~  \text{where} ~~ 
 \gbsigma = \Big(\frac{g_{\sigma}}{m_{\sigma}} \Big) ~,
\end{equation}
and $n_I^S = \langle\bar{\psi}_{I}\psi_{I}\rangle$ is the pseudo-scalar number 
density of the baryon. Similarly, by using the RMF approximation, the 
Euler-Lagrange equation for the temporal component of the $\omega$ meson,  
leads to
\begin{equation}\label{EoMOmega}
\omegab + \frac{\zeta \,\gbomega^4}{6} \, {\omegab}^{3} 
= \gbomega \sum_{I=n,p} n_I  ~,~~ \text{where} ~~ 
 \gbomega = \Big(\frac{g_{\omega}}{m_{\omega}} \Big) ~,
\end{equation}
where $\omegab = m_{\omega} \langle\hat{\omega}^0\rangle e^{\Phi}$ and 
$n_{I} = \langle\bar{\psi}_{I}\gamma^{0}\psi_{I}\rangle
= \langle\psi_{I}^{\dagger}\psi_{I}\rangle$ is the respective baryon number 
density. The eq. (\ref{EoMOmega}) can be solved exactly in terms of the total 
baryon number density $N \equiv \sum_{I=n,p} n_I$ as
\begin{equation}\label{OmegaSolution}
 \omegab = \frac{w_2^2 - 2\zeta}{\zeta \gbomega^2 w_2} ~,
\end{equation}
where $w_2 = (w_1 + \zeta^2 w_0)^{1/3}$ with $w_1 = \sqrt{\zeta^3 (8 + \zeta 
w_0^2)}$ and $w_0 = 3\, \gbomega^3 N$. From the eqs. (\ref{EoMSigma}, 
\ref{EoMOmega}) we note that the vacuum expectation values of the meson fields 
are dynamically determined through the number density $n_I$ and pseudo-scalar 
number density $n_I^S$.

Now by using the locally flat coordinates $(t,X,Y,Z)$ inside the box, together 
with the RMF approximation, the action for meson fields can be reduced to
\begin{equation}\label{MesonActionReduced}
S_M = \int d^{4}x \Big[\mathcal{L}_{\sigma\omega} + \sum_{I=n,p} \bar{\psi}_I 
 e^{\Phi} (\gbsigma \sigmab - \gamma^0 \gbomega \omegab) \psi_I \Big] ~,
\end{equation}
where
\begin{equation} \label{LagrangianSigmaOmega}
\mathcal{L}_{\sigma\omega} =  e^{\Phi}\left[
- \frac{ \sigmab^2}{2} + \frac{\omegab^2}{2} 
+ \frac{\zeta\, \gbomega^4}{24} \, \omegab^4 \right] ~.
\end{equation}
As earlier, the reduced action (\ref{MesonActionReduced}) can be viewed as a 
modified action written in the Minkowski spacetime. However it carries the
information about the global metric function $\Phi$. We use this modified 
action of mesons for subsequent analysis.

\subsection{Partition function}\label{subsec:PartitionFunction}

We follow the approach of thermal quantum field theory \cite{matsubara1955new, 
kapusta2006finite} to derive the equation of state for the $\sigma-\omega$ model 
(\ref{SigmaOmegaTotalAction}). The corresponding grand canonical partition 
function can be written as
\begin{equation}\label{GrandCanonicalPartitionFunction}
\mathcal{Z} = \text{Tr}[e^{-\beta(\hat{H} - \sum_{I}\mu_{I}\hat{n}_{I})}] ~,
\end{equation}
where $\mu_{I}$ is the chemical potential of the $\Ith$ constituent fermion with 
$\hat{n}_{I}$ being its number operator. Further, $\hat{H}$ is the Hamiltonian 
operator of the system and $\beta = 1 /(k_{B}T)$ with $k_{B}$ and $T$ being the 
Boltzmann constant and the temperature respectively. In the functional 
integral approach using the coherent states of fermions \cite{kapusta2006finite, 
Das:1997gg}, the partition function (\ref{GrandCanonicalPartitionFunction}) 
becomes
\begin{equation}\label{IntegralPartitionFunction}
\mathcal{Z} = \int\prod_{I}\mathcal{D}\bar{\psi}_{I}\mathcal{D}\psi_{I} 
e^{-S^{\beta}} ~,
\end{equation}
where the Euclidean action is defined as $S^{\beta} = \int_{0}^{\beta}d\tau\int 
d^{3}x (\mathcal{L}^{E} - \sum_{I} \mu_{I} \bar{\psi}_{I}\gamma^{0}\psi_{I})$ 
and the Euclidean Lagrangian density is obtained through the Wick rotation  
$\mathcal{L}^{E} = -\mathcal{L}(t\rightarrow -i\tau)$. In the expression (\ref{IntegralPartitionFunction}), the standard Minkowskian measure 
is used for spinor field, as here we have employed an effective action 
(\ref{FermionActionReduced}) approach. However, in a general approach 
for an arbitrary curved spacetime one may need to include appropriate factor 
of metric density in the path integral measure \cite{toms1987functional}. 
Here we have chosen the convention of the Wick rotation as in the ref. 
\cite{Das:1997gg, kapusta2006finite}, unlike the one used in the ref.  
\cite{hossain2021equation}. However, this choice does not change any result for 
non-interacting degenerate neutrons as considered in \cite{hossain2021equation}. 
We note that in an arbitrary curved spacetime with time-dependent metric, the 
Wick rotation as used here, cannot be always applied\cite{visser2017wick}. 
However, the interior spacetime of the neutron stars as studied here, is 
spherically symmetric hence static. Therefore, the said problem of Wick 
rotation does not affect the spacetime as studied here.
By using the reduced actions (\ref{FermionActionReduced}, 
\ref{MesonActionReduced}) one can split the partition function 
(\ref{IntegralPartitionFunction}) as
\begin{equation}\label{LnPartitionFunction}
\ln \mathcal{Z} = \beta V \mathcal{L}_{\sigma\omega} 
+ \sum_I \ln \mathcal{Z}_{\psi_I} ~,
\end{equation}
where $V$ is the volume of the box. The partition function involving $\Ith$ 
spinor can be expressed as $\mathcal{Z}_{\psi_I} = \int \mathcal{D} 
\bar{\psi}_{I}\mathcal{D}\psi_{I} \, e^{-S^{\beta}_{\psi_I}}$ where 
\begin{equation} \label{EuclideanFermionLagrangian}
S^{\beta}_{\psi_I} = \int_{0}^{\beta}d\tau\int d^{3}x \bar{\psi}_{I} 
\big[-\gamma^{0}(\partial_{\tau} + \mu_I^{*}) + e^{\Phi} ( 
i\gamma^{k}\partial_{k} + m_{I}^{*}) \big]\psi_{I} ~.
\end{equation}
In the eq. (\ref{EuclideanFermionLagrangian}), the \emph{effective} 
chemical potential is
\begin{equation}\label{EffectiveChemicalPotential}
\mu_I^{*} = \mu_I - \gbomega\omegab \, e^{\Phi} 
(\delta_{I}^{p} + \delta_{I}^{n})  ~,
\end{equation}
whereas the \emph{effective} mass is given by
\begin{equation}\label{EffectiveMass}
m_I^{*} = m_I - \gbsigma \sigmab \, (\delta_{I}^{p} + \delta_{I}^{n}) ~.
\end{equation}
We note that for an electron $\mu_{e}^{*} = \mu_{e}$ and $m_{e}^{*} = 
m_{e}$, given it has no coupling with the mesons.
In order to evaluate the partition function (\ref{LnPartitionFunction}), it is 
convenient to define the Fourier transformation of the spinor fields as
\begin{equation}\label{FourierTransform}
\psi_I(\tau,\x) = \frac{1}{\sqrt{V}} \sum_{l,\k} ~e^{-i(\omega_l\tau + 
\k\cdot\x)} \tilde{\psi}_I(l,\k)  ~,
\end{equation}
where $\omega_{l} = (2l+1) \pi \beta^{-1}$, with integers $l$, are the 
Matsubara frequencies for fermions arising due to the anti-periodic boundary 
condition $\psi_{I}(\tau+\beta,\x) =- \psi_{I}(\tau,\x)$ corresponding to the 
equilibrium temperature $T$. In the Fourier domain, the Euclidean action then 
becomes
\begin{equation}
S^{\beta}_{\psi_I} = \sum_{l,\k} \bar{\tilde{\psi}}_I(l,\k)
~ \beta \left[ \slashed{p} + \bar{m}_I \right] \tilde{\psi}_I(l,\k)  ~,
\end{equation}
where $\slashed{p} = \gamma^0(i \omega_l - \mu_I^{*}) + \gamma^{k}(\k_k 
e^{\Phi})$ and $\bar{m}_I = m_I^{*} e^{\Phi}$. A degenerate nuclear matter 
is characterized by the conditions $\beta \mu_I \gg 1$. Using these 
conditions, together with the results of integration over the Grassmann 
variables, one can evaluate the partition function for the $\Ith$ spinor as
\cite{hossain2021equation} 
\begin{equation}\label{IthPartionFunction}
\ln\mathcal{Z}_{\psi_I}  = \frac{\beta V e^{-3\Phi}}{24\pi^2} 
\left[ 2 \mu_I^{*} \mu_{Im}^3 - 3 \bar{m}_{I}^2 \, \muimb^2 \right] ~,
\end{equation}
where we have defined $\mu_{Im} \equiv \sqrt{\mu_I^{*2} -\bar{m}_{I}^2}$ and
$\muimb^2 \equiv \mu_I^{*} {\mu}_{Im} - \bar{m}_{I}^2 \, \arcsinh 
(\mu_{Im}/\bar{m}_I)$. In the expression (\ref{IthPartionFunction}), we have 
neglected the finite-temperature correction terms, as these corrections terms 
are very small for the system under consideration.

\subsection{Pressure and energy density}

The number density of $\Ith$ spinor can be computed using the expression 
$n_I = (\beta V)^{-1} (\partial \ln \mathcal{Z} /\partial \mu_I)$ whereas its 
pseudo-scalar number density can be computed as $e^{\Phi} n^S_I = - (\beta 
V)^{-1} (\partial \ln \mathcal{Z} /\partial m_I)$. The evaluation of these 
partial derivatives by using the eqs. (\ref{LnPartitionFunction}, 
\ref{IthPartionFunction}), leads to
\begin{equation}\label{IthNumberDensity}
 n_I   = \frac{ e^{-3\Phi}}{3\pi^2} ~ \mu_{Im}^3 ~~,~~
 n^S_I = \frac{e^{-3\Phi}}{2\pi^2} ~ \bar{m}_I \muimb^2 ~,
\end{equation}
where we have used the relations $(\partial \muimb^2/\partial\mu_I^{*}) = 
2\mu_{Im}$, $(\partial\mu_{Im}/\partial\mu_I^{*}) = (\mu_I^{*}/\mu_{Im})$,
$(\partial\mu_{Im}/\partial \bar{m}_I) =  - (\bar{m}_I/\mu_{Im})$,
and $(\partial \muimb^2/\partial \bar{m}_I) = - 2 \bar{m}_I \, \arcsinh 
(\mu_I^{*}/\bar{m}_I) $.
For a grand canonical ensemble, total pressure $P = (\beta V)^{-1}\ln 
\mathcal{Z}$, becomes
\begin{equation}\label{CurvedPressure}
P =  P_{\sigma\omega} +  \sum_{I} P_{I} ~,
\end{equation}
where the term involving only meson contributions is
\begin{equation} \label{LnPFSigmaOmega}
P_{\sigma\omega} = e^{\Phi}\left[
- \frac{ \sigmab^2}{2} + \frac{\omegab^2}{2} 
+ \frac{\zeta\, \gbomega^4}{24} \, \omegab^4 \right] ~.
\end{equation}
On the other hand, the pressure contribution involving $\Ith$ spinor is 
given by
\begin{eqnarray}\label{IthPressureCurved}
P_I = \frac{e^{\Phi} m_I^{*4}}{24\pi^2} \left[ \sqrt{(b_I n_I)^{\frac{2}{3}} 
+ 1} \left\{2(b_I n_I) - 3(b_I n_I)^{\frac{1}{3}} \right\} \nonumber \right. \\
+ \left. ~3 \arcsinh \left\{ (b_I n_I)^{\frac{1}{3}}\right\} \right] ~,~~~
\end{eqnarray}
where the constant $b_I = 3\pi^2/m_I^{*3}$. Using the partition function
(\ref{LnPartitionFunction}), the energy density $\rho$ within the 
box can be computed as $(\rho - \sum_I \mu_I n_I)V = -(\partial \ln 
\mathcal{Z} /\partial \beta)$ which leads to
\begin{equation}\label{CurvedEnergyDensity}
\rho =  \rho_{\sigma\omega} + \sum_{I} \rho_{I} ~.
\end{equation}
In the eq. (\ref{CurvedEnergyDensity}), the direct meson contributions are
\begin{equation} \label{EnergyDensitySigmaOmega}
\rho_{\sigma\omega} =  e^{\Phi}\left[
 \frac{ \sigmab^2}{2} + \frac{\omegab^2}{2} 
+ \frac{\zeta\, \gbomega^4}{8} \, \omegab^4  \right] ~,
\end{equation}
whereas the contribution due to the $\Ith$ spinor is 
\begin{equation}\label{IthEnergyDensityCurved}
\rho_I = - P_I + \frac{e^{\Phi} m_I^{*4}}{3\pi^2} 
\, (b_I n_I) \, \sqrt{(b_I n_I)^{\frac{2}{3}} + 1}  ~.
\end{equation}
As one expects, the expressions for pressure $P$ (\ref{CurvedPressure}) and 
energy density $\rho$ (\ref{CurvedEnergyDensity}), which we shall refer to as 
the \emph{curved} EOS, reduce exactly to their flat spacetime counterparts
\emph{i.e.} the \emph{flat} EOS, when one sets the metric function 
$\Phi = 0$. This imposition is equivalent of setting the lapse function 
$e^{\Phi}=1$ in the entire interior spacetime of the star. Clearly, the equation 
of states which are used in solving the TOV eqs. but computed in the Minkowski 
spacetime, fail to capture the effect of general relativistic time dilation.

\subsection{Number density relations in $\beta$-equilibrium}

By using the eq. (\ref{IthNumberDensity}), the effective chemical potentials 
can be expressed in terms of the number densities as
\begin{equation}\label{MuINIRelation}
e^{-\Phi} \mu_I^{*} = m_I^{*} \sqrt{(b_I n_I)^{\frac{2}{3}} + 1 } ~,
\end{equation}
whereas the expression of the pseudo-scalar number density can be written as
\begin{equation}\label{PseudoScalarNumberDensity}
n_I^S = \frac{3}{2 b_I} \big[(b_I n_I)^{\frac{1}{3}} 
\sqrt{(b_I n_I)^{\frac{2}{3}} + 1} 
- \arcsinh\big\{(b_I n_I)^{\frac{1}{3}}\big\} \big] .
\end{equation}
Nevertheless, these number densities, corresponding to the three fermions, 
namely $n \equiv n_n$, $n_p$ and $n_e$ are not independent and are subject to 
the conditions
\begin{equation}\label{NumberDensityConstraints}
n_p = n_e ~~\text{and}~~ \mu_n = \mu_p + \mu_e ~.
\end{equation}
In the eqs. (\ref{NumberDensityConstraints}), the first equality follows from
the $U(1)$ charge balance condition, as the electron is the only lepton here. 
The second equality follows from the $\beta$-equilibrium reaction $n 
\leftrightarrow p + e$ which in turn imposes a condition on the chemical 
potentials. Therefore, we can independently choose only \emph{one} out of 
these three number densities.

\section{Numerical evaluation}\label{sec:NumericalEvaluation}

The action (\ref{SigmaOmegaTotalAction}) contains apriori \emph{five} free 
parameters $m_{\sigma}$, $g_{\sigma}$, $m_{\omega}$, $g_{\omega}$ and $\zeta$ 
that govern the dynamics of the meson fields $\sigma$ and $\omega$. 
However, the equations of motion, as described by the eqs. (\ref{EoMSigma}, 
\ref{EoMOmega}), depend only through the ratios $g_{\sigma}$ to $m_{\sigma}$ 
and $g_{\omega}$ to $m_{\omega}$. Consequently, there are only 
$\emph{three}$ free parameters, namely $\gbsigma$, $\gbomega$ and $\zeta$, in 
the equation of state (\ref{CurvedPressure}, \ref{CurvedEnergyDensity}) 
of the $\sigma-\omega$ model. For convenience of numerical evaluation, we define 
following \emph{dimensionless} quantities representing the scaled coupling 
constants of mesons as
\begin{equation}\label{DimensionlessCouplings}
 \gtsigma = \gbsigma m_n ~,~ \gtomega = \gbomega m_n ~,
\end{equation}
where $m_n$ is the bare mass of a neutron and we set its value to 
be $939.57$ MeV.

In the literature, there exist several versions of the $\sigma-\omega$ 
model having a varied range of parameter values as summarized in the ref.
\cite{lim2018effective}. In particular, the values of the coupling constants 
$g_{\sigma}$ range between $7.5-10.2$, $g_{\omega}$ range between 
$8.7-12.9$, and  $\zeta$ range between $0 - 0.06$. On the other hand, 
the values of the masses, $m_{\sigma}$ range between $467.0-507.3$ MeV and 
$m_{\omega}$ range between $761.2-781.3$ MeV. These values together then imply 
that the range of $\gtsigma$ to be $13.9-20.5$ whereas the range of $\gtomega$ 
to be  $10.5-15.9$. In the literature, apart from the fields that are considered 
here, the hyperons, $\rho$ and $\phi$ mesons are also included often in the 
$\sigma-\omega$ model. However, for simplicity, here we have not included 
these fields. Besides, integrating out the hyperon degrees of freedom changes 
the masses of mesons which we could change independently in the numerical 
evaluation. Secondly, the $\rho$ and $\phi$ mesons are coupled to baryons 
via Yukawa interaction. So integrating out these mesons generates an 
effective four baryon interaction which changes the effective masses of the 
baryons.

In order to keep the focus on the effect of gravitational time dilation here we 
treat the EOS corresponding to $\sigma-\omega$ model to remain valid for the 
entire range of baryon number density during the numerical evaluation. In 
particular, we do not interpolate to a different equation of state at a low 
baryon number density. It also keeps the analysis much simpler.

\subsection{Particle fractions and effective mass}

\begin{figure}
\includegraphics[height=5.9cm, width=7cm]{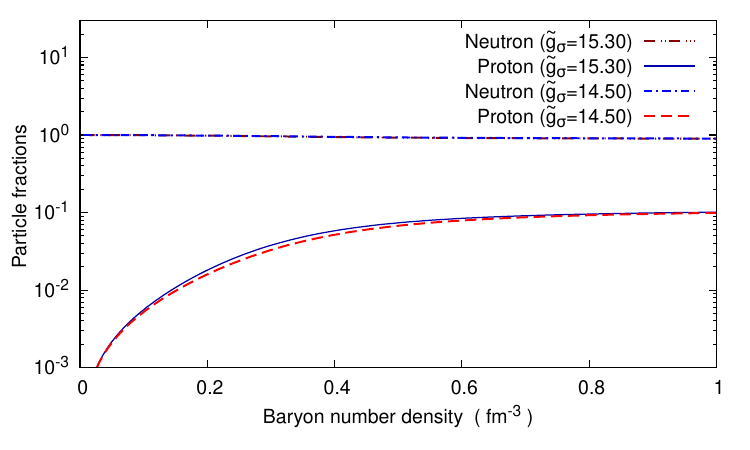}
\caption{Particle fractions as a function of baryon number density $N$ for 
different values of $\gtsigma$.}
\label{fig:ParticleFraction}
\end{figure}

For the purpose of numerical evaluation we consider the bare masses of protons 
and neutrons to be equal \emph{i.e.} $m_n = m_p$. Given $m_n \gg m_e$, the 
eqs. (\ref{MuINIRelation}, \ref{NumberDensityConstraints}) lead to 
\begin{equation}\label{ProtonNumberDensity}
n_p = n_e \simeq \frac{(b_n n)^2} {8 b_n (\sqrt{(b_n n)^{2/3} + 1}~)^3} ~.
\end{equation}
The eq. (\ref{ProtonNumberDensity}) should be regarded as a function $n_p 
= n_p(n,\sigmab)$ whereas the eqs. (\ref{EoMSigma}, \ref{IthNumberDensity}) 
together imply $\sigmab = \sigmab(n,n_p)$. So to find the number density of 
proton, one must solve for both $\sigmab$ and $n_p$ simultaneously by using the 
eqs. (\ref{EoMSigma}, \ref{IthNumberDensity}, \ref{ProtonNumberDensity}).
Here we use numerical root finding methods to find the solution $n_p$ and 
$\bar{\sigma}$ for a given neutron number density $n$. The numerically 
evaluated particle fractions for neutrons and protons, as a function of baryon 
number density $N = n + n_p$, are plotted in the FIG. 
\ref{fig:ParticleFraction}. It may be noted that a higher value of $\gtsigma$ 
leads to a higher proton fraction.

\begin{figure}
\includegraphics[height=5.9cm, width=7cm]{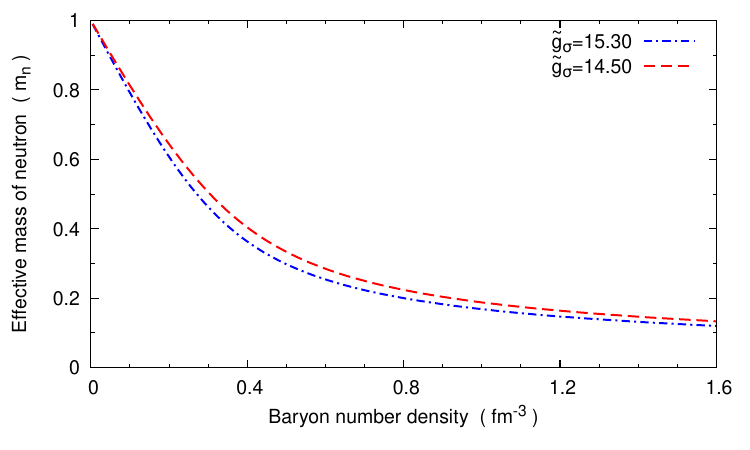}
\caption{Effective mass of the neutron as a function of baryon number 
density $N$ for different values of coupling constant $\gtsigma$.}
\label{fig:EffectiveMass}
\end{figure}

In the FIG. \ref{fig:EffectiveMass}, the effective mass of the neutron is 
plotted for different values of the scaled coupling constant $\gtsigma$. It 
may be observed that an increase of the parameter $\gtsigma$ leads the 
effective mass to decrease.

\subsection{Kinematical behavior of curved EOS}

\begin{figure}
\includegraphics[height=5.9cm, width=7cm]{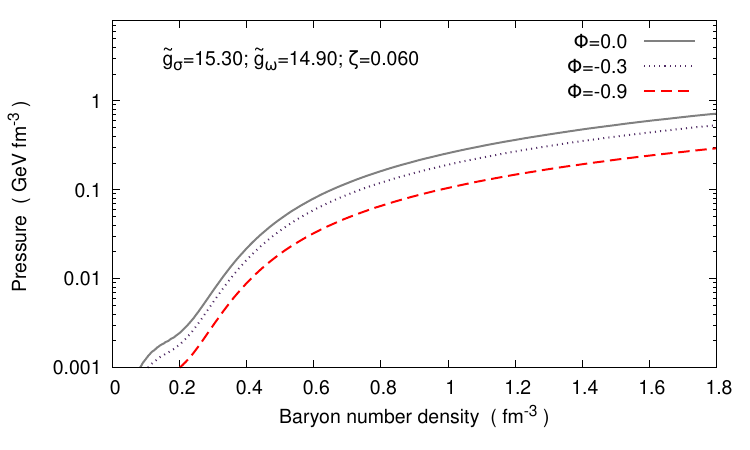}
\caption{Plot of the pressure $P$ as a function of baryon number density $N$ for 
different kinematical values of $\Phi$.}
\label{fig:PressureComparison}
\end{figure}

\begin{figure}
\includegraphics[height=5.9cm, width=7cm]{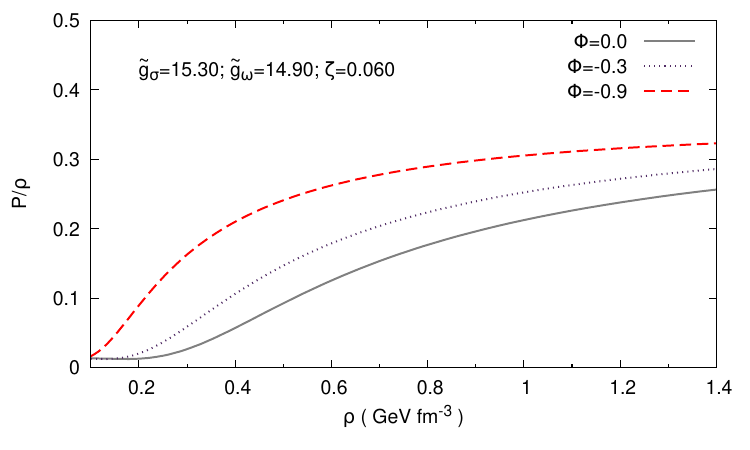}
\caption{Ratio between the pressure and the energy density, $(P/\rho)$, is 
plotted as a function of $\rho$ for different kinematical values of the metric 
function $\Phi$.}
\label{fig:EOSComparison}
\end{figure}

For different kinematical values of the metric function $\Phi$, pressure of 
the curved EOS (\ref{CurvedPressure}) is plotted as a function of baryon 
number density $N$ in the FIG. \ref{fig:PressureComparison}. We note that for 
a given baryon number density, presence of the metric function $\Phi$ in 
the curved EOS leads to a reduction of the pressure, compared to its flat 
spacetime counterpart ($\Phi=0$). Similar behavior is seen also in the 
expression of the energy density (\ref{CurvedEnergyDensity}). 
Consequently, the presence of $e^{\Phi}$ factor in the expressions 
(\ref{CurvedPressure}, \ref{CurvedEnergyDensity}) makes the pressure $P$ of the 
curved EOS comparatively \emph{stiffer} for higher values of the energy 
density $\rho$. This behavior is shown in the FIG. \ref{fig:EOSComparison} 
for different kinematical values of the metric function $\Phi$.

We may emphasize that here we have chosen a set of fixed kinematical 
values of $\Phi$ while plotting the figures given in FIG. 
\ref{fig:PressureComparison} and FIG. \ref{fig:EOSComparison}. 
However as follows from the TOV eqs. (\ref{TOVEqn}), the metric 
function $\Phi$ varies dynamically along the radial direction inside the stars.

\subsection{Numerical method for solving TOV eqs.}

We note from the eq. (\ref{ProtonNumberDensity}) that number densities of 
the fermions, after solving for $\sigmab$, can be viewed as functions of 
neutron number density $n$ as $n_I = n_I(n)$.  Therefore, the pressure 
(\ref{CurvedPressure}) and the energy density (\ref{CurvedEnergyDensity}) can be 
viewed as explicit functions of $n$ and $\Phi$, given as $P(n,\Phi)$ and 
$\rho(n,\Phi)$ respectively. Consequently, we can treat the TOV eqs. 
(\ref{TOVEqn}) as a set of first-order differential equations for the 
\emph{triplet} $\{\mr,\Phi,n\}$ where the neutron number density $n$ satisfies
\begin{equation}\label{NeutronNumberEqn}
\frac{dn}{dr} = - \frac{(\rho + P + ({\partial P}/{\partial \Phi}))}
{({\partial P}/{\partial n})} \frac{d\Phi}{dr} ~.
\end{equation}
On the other hand, for the flat EOS, the pressure and the energy density are 
\emph{independent} of $\Phi$ . Consequently, $\Phi$ can be eliminated from the 
set of TOV eqs. (\ref{TOVEqn}) which then can be viewed as a set of first-order 
differential equations for the \emph{doublet} $\{\mr,n\}$.

For the curved EOS, nevertheless, the triplet $\{\mr,\Phi,n\}$ is subject to 
the boundary conditions $e^{2\Phi(R)} = (1 - 2G M /R)$, $M = \mr(R)$ and $n(R) = 
0$ \emph{i.e.} interior metric of a star of mass $M$ and radius $R$ must match 
with the Schwarzschild metric at the surface.
In order to evolve the eq. (\ref{NeutronNumberEqn}) numerically, apart from 
$\rho$ and $P$, we also need to compute the terms $({\partial P}/{\partial 
\Phi})$ and $({\partial P}/{\partial n})$. In particular, the term $({\partial 
P}/{\partial n})$ can be expressed as
\begin{equation}\label{PressureDerivative}
\frac{\partial P}{\partial n}  = 
\sum_{I=n,p,e} \frac{\partial P_I}{\partial n_I} \frac{dn_I}{dn}
+ \frac{\partial P}{\partial \omegab} \frac{d \omegab}{d n} 
+ \frac{\partial P}{\partial \sigmab} \frac{d \sigmab}{d n} ~,
\end{equation}
where partial derivatives of $P$ \emph{w.r.t.} $n_I$, $\omegab$ are
\begin{equation}
\frac{\partial P_{I}}{\partial n_{I}} = 
\frac{e^{\Phi} m_{I}^{*} (b_I n_I)^{2/3}} {3\sqrt{(b_I n_I)^{2/3} + 1} } ~~,~
\frac{\partial P}{\partial\omegab} = e^{\Phi}\Big[\omegab + \frac{\zeta}{6}
\gbomega^{4}\omegab^{3}\Big] ~,
\end{equation}
respectively and partial derivative of $P$ \emph{w.r.t.} $\sigmab$ is
\begin{equation}
\frac{\partial P}{\partial\sigmab} = - e^{\Phi}\sigmab + 3\gbsigma 
\sum_{I=n,p} \frac{n_I}{m_I^{*}} 
\left(\frac{\partial P_I}{\partial n_I} - \frac{4 P_I}{3 n_I}\right) ~.
\end{equation}
Total derivative of $\omegab$ (\ref{OmegaSolution}) \emph{w.r.t.} $n$ can 
be expressed as
\begin{equation}
\frac{d\omegab}{dn} = \frac{\zeta \gbomega (2\zeta + w_2^2)}{w_1 w_{2}}
 \sum_{I=n,p} \frac{d n_I}{d n} ~.
\end{equation}
On the other hand, total derivatives of the number densities $n_I$ 
can be expressed as
\begin{equation}\label{nITotalDerivative}
\frac{d n_I}{d n}  = \frac{\partial n_I}{\partial n}
+ \frac{\partial n_I}{\partial \sigmab} \frac{d \sigmab}{d n} ~,
\end{equation}
where partial derivatives \emph{w.r.t.} $n$ are
\begin{equation}\label{DelnIDeln}
\frac{\partial n_p}{\partial n} = \frac{\partial n_e}{\partial n} 
= \frac{(b_n n)}{8} \frac{(b_n n)^{2/3} + 2}{(\sqrt{(b_n n)^{2/3} + 1})^5} ~,
\frac{\partial n}{\partial n} = 1 ~,
\end{equation}
and partial derivatives \emph{w.r.t.} $\sigmab$ are given by
\begin{equation}\label{DelnIDelsigmab}
\frac{\partial n_p}{\partial\sigmab} = \frac{\partial n_e}{\partial\sigmab} 
= \frac{\gbsigma m_n^{*2}}{8\pi^2} 
\frac{(b_n n)^2}{(\sqrt{(b_n n)^{2/3} + 1})^5} ~,
\frac{\partial n}{\partial\sigmab} = 0 ~.
\end{equation}
We have discussed earlier that one needs to solve for $\sigmab$ for a given 
neutron number density $n$. Once $\sigmab$ is found, the total derivative of 
$\sigmab$ can be expressed as
\begin{equation}\label{dsigmabdn}
\frac{d\sigmab}{dn} = \gbsigma
\Big[\sum_{n,p}\frac{\partial n_I^S}{\partial n_I}
\frac{\partial n_I}{\partial n} \Big]
\Big[1 - \gbsigma\sum_{n,p} \Big(\frac{\partial n_{I}^{S}}{\partial\sigmab} 
+ \frac{\partial n_{I}^{S}} {\partial n_{I}}\frac{\partial 
n_{I}}{\partial\sigmab}\Big)\Big]^{-1}  ,
\end{equation}
where partial derivatives of $n_I^S$ with \emph{w.r.t.} $n$ and $\sigmab$ 
can be expressed as
\begin{equation}\label{DelsnIDelnI}
\frac{\partial n_I^S}{\partial n_I} = \frac{1}{\sqrt{(b_I n_I)^{2/3} + 1}} ~, 
\end{equation}
and 
\begin{equation}\label{DelsnIDelsigmab}
\frac{\partial n_I^S}{\partial\sigmab} = - \frac{\gbsigma m_I^{*2}}{\pi^2}
\Big[(b_I n_I^S) - \frac{(b_I n_I)}{\sqrt{(b_I n_I)^{2/3} + 1}} \Big] ~.
\end{equation}
The eqs. (\ref{DelnIDeln}, \ref{DelnIDelsigmab}, \ref{DelsnIDelnI}, 
\ref{DelsnIDelsigmab}) completely determine the value of $d\sigmab/dn$ for a 
given value $n$ and $\sigmab$.

In summary, the TOV eqs. (\ref{TOVEqn}) here can be viewed as a well-defined 
boundary value problem. In order to satisfy the boundary condition 
numerically, we begin with a given central neutron number density, say $n_c$ and 
a trial value of $\Phi$ at the center. Subsequently, we evolve the TOV eqs. 
towards the surface by computing the meson field values $\sigmab$ and $\omegab$ 
at each step. This leads to an evolved value of $\Phi$ at the surface which is 
then compared with an independently calculated value of $\Phi$ at the surface by 
using the boundary condition, say, $\Phi_s = \tfrac{1}{2}\ln(1 - 2G M /R)$. In 
the next step, $\Phi$ at the center is numerically computed starting from 
the value $\Phi_s$ by evolving backward from $n=0$ at the surface to $n=n_c$ at 
the center by using the eq. 
\begin{equation}\label{dndPhiEqn}
\frac{d\Phi}{dn} = - \frac{({\partial P}/{\partial n})} {(\rho + P + 
({\partial P}/{\partial \Phi}))} ~,
\end{equation}
which follows from the eq. (\ref{NeutronNumberEqn}). These steps are iterated in 
order to achieve the convergence between the evolved and the computed values of 
the metric function $\Phi$ at the center within the desired numerical precision. 
The said iteration method converges rapidly except for the situations where the 
mass $M$ of the neutron star changes without  appreciable change of the radius 
$R$. For these situations, one needs to employ an appropriate root finding 
method.

\subsection{Mass-radius relations}

\begin{figure}
\includegraphics[height=5.9cm, width=7cm]{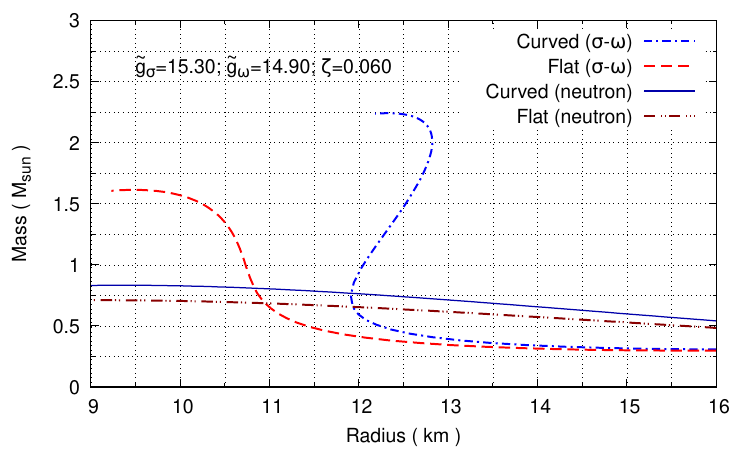}
\caption{Comparison of mass-radius relations of neutron stars whose 
nuclear matters are described by the $\sigma-\omega$ model and an ensemble of 
non-interacting degenerate neutrons.}
\label{fig:MRComparisonNSOIDN}
\end{figure}

\begin{figure}
\includegraphics[height=5.9cm, width=7cm]{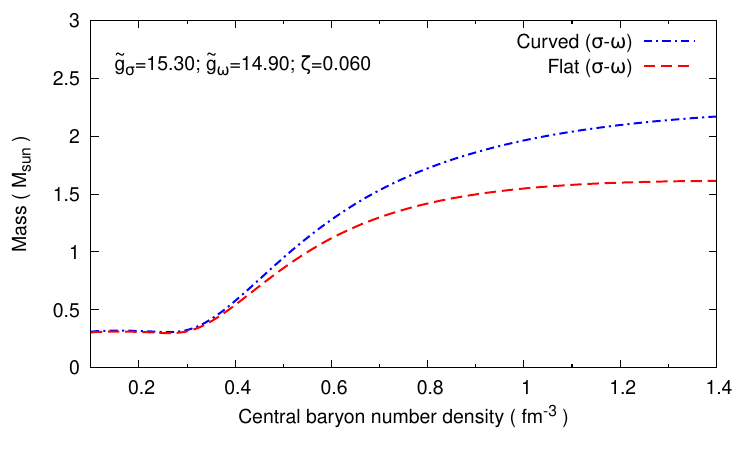}
\caption{Plot of the neutron star mass $M$ as a function of the
central baryon number density.}
\label{fig:MassVsCentralDensity}
\end{figure}

In the FIG. \ref{fig:MRComparisonNSOIDN}, we compare the mass-radius relations 
arising from both the curved EOS and the flat EOS for the $\sigma-\omega$ model 
and an ensemble of ideal non-interacting degenerate neutrons. It can be seen 
that irrespective of how the nuclear matters are described, the usage of the 
curved EOS, rather than the flat EOS, leads to a \emph{significantly} higher 
mass limit.
In the FIG. \ref{fig:MassVsCentralDensity}, we plot the dependency of 
neutron star mass $M$ on the central baryon number density for both the curved 
EOS and the flat EOS.

\begin{figure}
\includegraphics[height=5.9cm, width=7cm]{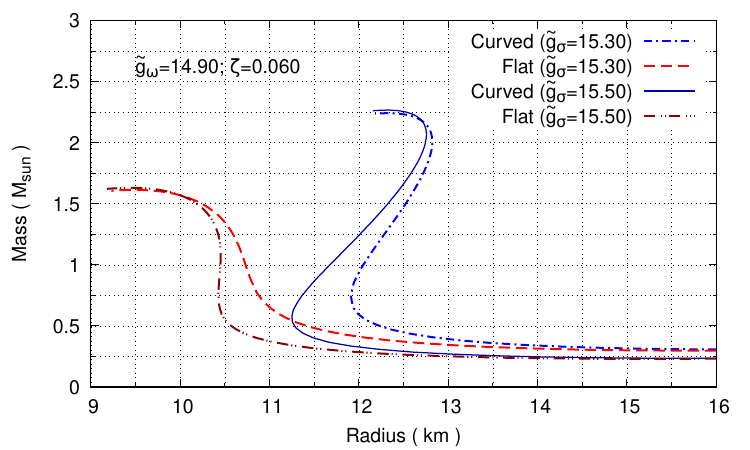}
\caption{Plot of the mass-radius relations for different values of the parameter
$\gtsigma$.}
\label{fig:MRSigma}
\end{figure}

We have mentioned earlier that the equation of state corresponding to the 
$\sigma-\omega$ model, contains three independent parameters, 
namely $\gtsigma$, $\gtomega$, and $\zeta$. The parameter $\gtsigma$ changes 
the nature of the turning point of the mass-radius curve. An increase in 
the value of $\gtsigma$ changes the position of the turning point towards a 
smaller radius. Moreover, as the value of $\gtsigma$ increases, the radius of 
the neutron star with a given mass decreases and the maximum mass of the neutron 
star increases. The dependency of mass-radius relations on the parameter 
$\gtsigma$ is shown in the FIG. \ref{fig:MRSigma}.

\begin{figure}
\includegraphics[height=5.9cm, width=7cm]{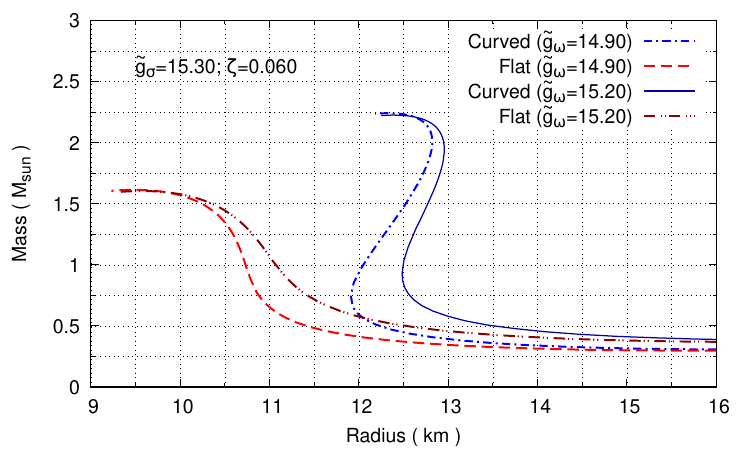}
\caption{Plot of the mass-radius relations for different values of the parameter
$\gtomega$.}
\label{fig:MROmega}
\end{figure}

\begin{figure}
\includegraphics[height=5.9cm, width=7cm]{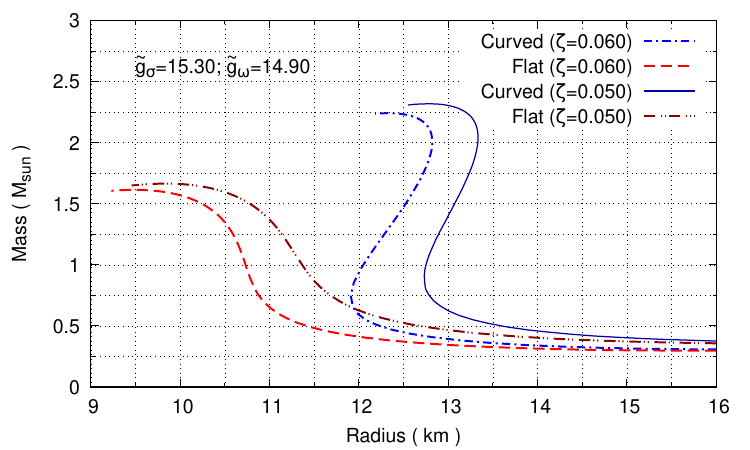}
\caption{Plot of the mass-radius relations for different values of the 
parameter $\zeta$.}
\label{fig:MRZeta}
\end{figure}

The parameter $\gtomega$ also changes the nature of the turning point of 
mass-radius curve. An increase in the value of $\gtomega$ changes position of 
the turning point towards a larger radius and after a certain value of 
$\gtomega$, the turning point disappears altogether. As the value of $\gtomega$ 
increases, radius of the neutron star with a given mass increases. Moreover, as 
the value of $\gtomega$ increases, the maximum mass of neutron star decreases. 
This behavior is shown in the FIG. \ref{fig:MROmega}.
On the other hand, the parameter $\zeta$ changes the maximum mass and the 
corresponding radius without much alteration in the nature of mass-radius curve. 
An increase in the value of $\zeta$ causes the maximum mass limit of the neutron 
stars to decrease. Moreover, as the value of $\zeta$ increases, the radius of 
neutron star for a given mass decreases. This aspect is shown in the FIG. 
\ref{fig:MRZeta}.

Clearly, in all these mass-radius relations, a significant enhancement of the 
maximum mass limits can be seen when one uses the curved EOS rather than the 
flat EOS. A quantitative comparison of the computed mass limits and the 
corresponding radii of the neutron stars are given in the TABLE 
\ref{table:ParameterTable}. As an example, it can be seen that the flat EOS with 
the parameter values $\gtsigma = 15.30$, $\gtomega=14.90$, and $\zeta = 0.06$ 
leads the maximum mass limit to be around $1.61 M_{\odot}$ with radius of 
$9.50$ km. On the other hand, the curved EOS with the same set of parameters 
leads the maximum mass limit to be around $2.24 M_{\odot}$ with a radius of 
around $12.33$ km. Thus incorporation of the effect of gravitational time 
dilation enhances the maximum mass limit here by almost $39.1\%$. The 
corresponding increase in radius of the star is around $29.8\%$. 
These enhancements of mass limits are controlled quantitatively 
by the ratio $(GM/R)$ of the star and follow from the equation of state 
(\ref{CurvedPressure}, \ref{CurvedEnergyDensity}) as $\rho_{curved}(r + 
\Delta r) > \rho_{flat}(r + \Delta r)$ even if $\rho_{curved}(r) = 
\rho_{flat}(r)$ and $(\tfrac{\partial \rho_{curved}}{\partial n} \tfrac{d 
n}{dr})_{|r} = (\tfrac{\partial \rho_{flat}}{\partial n} \tfrac{d n}{dr})_{|r}$ 
given $(d\Phi/dr) > 0$ for $\Delta r > 0$ \cite{hossain2021equation}.

\begin{table}
\caption{The maximum mass limits and the corresponding radii of neutron 
stars for both \emph{curved} EOS and \emph{flat} EOS of the 
$\sigma-\omega$ model for different parameter sets.} 
\begin{tabular}{c c c c c c c}
\hline
~~~$\gtsigma$~~~ & ~~~$\gtomega$~~~ & ~~~$\zeta$~~~ ~~
& \multicolumn{2}{c}{~$M ~(M_{\odot})$~} 
& \multicolumn{2}{c}{~$R ~(km)$~} \\
~ & ~ & ~ & ~flat~ & ~curved~ & ~flat~ & ~curved~  \\
\hline
15.30 & 14.90 & 0.060 & 1.61 & 2.24 & 9.50 & 12.33  \\
15.30 & 15.20 & 0.060 & 1.60 & 2.23 & 9.59 & 12.40  \\
15.30 & 14.90 & 0.050 & 1.66 & 2.32 & 9.84 & 12.79  \\
15.50 & 14.90 & 0.060 & 1.63 & 2.27 & 9.45 & 12.30  \\
14.50 & 14.70 & 0.060 & 1.56 & 2.16 & 9.63 & 12.35  \\
15.90 & 14.90 & 0.020 & 2.00 & 2.85 & 11.24 & 14.87  \\
\hline 
\label{table:ParameterTable}
\end{tabular}
\end{table}

\begin{figure}
\includegraphics[height=5.9cm, width=7cm]{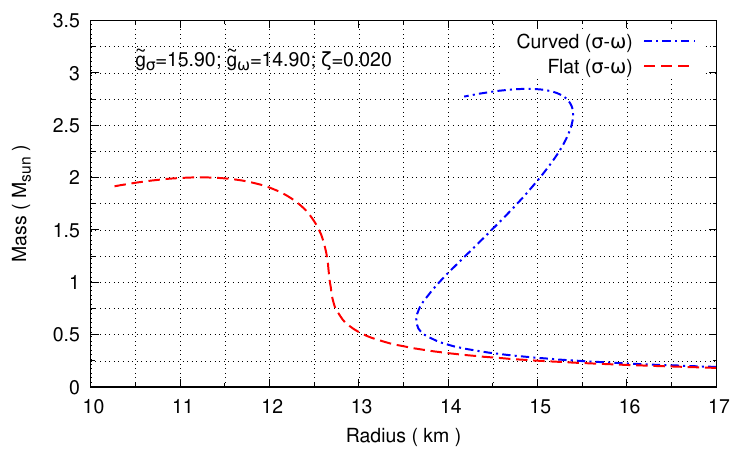}
\caption{Comparison of mass-radius relations of neutron stars whose 
nuclear matters are described by the $\sigma-\omega$ model with a chosen set 
of parameters such that both flat EOS and curved EOS leads the mass limits 
to be $2 M_{\odot}$ or higher.}
\label{fig:MRComparisonNSO2}
\end{figure}

We would like to note here that for another chosen set of parameters, 
the flat EOS corresponding to the $\sigma-\omega$ model leads to a maximum 
mass limit of around $2 M_{\odot}$. On the other hand, the corresponding 
curved EOS with the same set of parameters leads the maximum mass limit of  
neutron stars to be around $2.85 M_{\odot}$. These two mass-radius curves 
are plotted in the FIG. \ref{fig:MRComparisonNSO2}.

\section{Universal effect of gravitational time dilation}
\label{sec:UniversalityTimeDilation}

We have seen that the usage of flat EOS fails to capture the effect of 
gravitational time dilation. This aspect can be understood in a rather simple 
way. While solving the TOV eqs. (\ref{TOVEqn}), one evolves from the center to 
the surface of the star. In this process, the metric function $\Phi$ changes 
considerably. Consequently, the clock speed in a locally flat spacetime near the 
center of the star differs from the clock speed of a locally flat spacetime near 
the surface of the star, given these two frames have different lapse functions, 
$e^{\Phi(r=0)}$ and $e^{\Phi(r=R)}$ respectively. An equation of state which is 
computed in a globally flat spacetime fails to capture this varying nature of 
the lapse function or the resultant varying clock speed. On the other hand, the 
curved EOS, as studied here, incorporates the gravitational time dilation 
through the presence of metric function $\Phi$ in the expressions of 
pressure and energy density (\ref{CurvedPressure}, \ref{CurvedEnergyDensity}).

Nevertheless, it is shown in \cite{hossain2021equation} that it is possible to 
obtain the equation of state for a spherically symmetric curved spacetime 
starting from its flat spacetime counterpart without going through a 
first-principle derivation. In particular, to obtain the partition function in 
a spherically symmetric spacetime, one needs to use the following 
transformations \cite{hossain2021equation} 
\begin{equation}\label{FlatToCurvedTransformationRules}
\beta\rightarrow\beta e^{\Phi} ~,~~ \mu_{I}\rightarrow\mu_{I}e^{-\Phi} ~. 
\end{equation}
The transformation rules (\ref{FlatToCurvedTransformationRules})can be 
understood as follows. By re-defining the time coordinate $t \to \tilde{t} = 
e^{\Phi} t$, one can transform the metric (\ref{MetricInTOVBox}) within the 
box to be the standard Minkowski metric. Consequently, at thermal equilibrium, 
the anti-periodic boundary condition for the spinor fields, as employed in the 
eq. (\ref{FourierTransform}), then leads to the transformation rules (\ref{FlatToCurvedTransformationRules}).
The universality of the transformations (\ref{FlatToCurvedTransformationRules}) 
can be checked from the general expression of the partition function in the 
Minkowski spacetime which can always be written as
\begin{equation}\label{GeneralLnZFlat}
\ln\mathcal{Z}_{\text{flat}}[\beta, \{\mu_{I}\}, \{\lambda_{i}\}] 
= \beta^{-3}Vf(\{ \beta\mu_{I}\},\{\beta^{d_{i}}\lambda_{i}\}) ~,
\end{equation}
where $\{\lambda_{i}\}$ is a set of parameters of an interacting 
matter field theory, having canonical mass dimensions $\{d_{i}\}$. For 
example, in the $\sigma-\omega$ model that we have studied here, the canonical 
mass dimensions of the coupling constants $\{g_{\sigma}, g_{\omega}, \zeta\}$ 
are \emph{zero} whereas the canonical mass dimensions of $\omega$ and $\sigma$ 
fields are \emph{one}. The general form (\ref{GeneralLnZFlat}) follows from the 
fact that $\ln\mathcal{Z}$ is a dimensionless, \emph{extensive} quantity in 
statistical physics. Therefore, by following the transformation rules 
(\ref{FlatToCurvedTransformationRules}), we can obtain the partition function 
in a spherically symmetric curved spacetime as
\begin{equation}\label{LnZRelation}
 \ln\mathcal{Z}[\beta,\{\mu_{I}\},\{\lambda_{i}\}]
 =  e^{-3\Phi} \ln\mathcal{Z}_{\text{flat}}[\beta,\{\mu_{I}\},
 \{e^{d_i\Phi}\lambda_i \}] ~.
\end{equation}
The eq. (\ref{LnZRelation}) can be used to obtain the partition function of 
the $\sigma-\omega$ model (\ref{LnPartitionFunction}) starting from its flat 
spacetime counterpart. It also shows that the different choices of frames for intermediate computation eventually lead to the same equation of state for 
the curved spacetime.

\section{Discussions}\label{sec:Discussions}

In summary, by employing a first-principle approach, we have derived the 
equation of state for a degenerate nuclear matter which is described by a 
simplified $\sigma-\omega$ model. Importantly, in this derivation the 
nuclear matter is assumed to reside within the spherically symmetric interior 
curved spacetime of the neutron star, rather than in the Minkowski spacetime as 
routinely used in the literature. The equation of state which is computed 
in the curved spacetime, includes the effect of gravitational time dilation. 
Furthermore, we have shown that the incorporation of gravitational 
time dilation significantly increases the maximum mass limits of neutron 
stars. As an example, the $\sigma-\omega$ model with a chosen set of parameters, 
leads the maximum mass limit to be around $1.61 M_{\odot}$ when one uses the 
equation of state computed in the Minkowski spacetime. In contrast, with the 
same set of parameters, the equation of state computed in the curved spacetime, 
leads the maximum mass limit to be around $2.24 M_{\odot}$, a significant 
increase of $\sim 39.1\%$.

Recent observations of several neutron stars having masses more than $2 
M_{\odot}$, have pushed many existing models of nuclear matters within the 
neutron stars, to be ruled out \cite{linares2018peering, 
cromartie2020relativistic}. However, as we have shown here that a proper 
incorporation of gravitational time dilation into the corresponding equation of 
states would enhance the maximum mass limits of such models.

Finally, we would like to emphasize here that the existence of the gravitational 
time dilation is a universal feature of the curved spacetime. Therefore, the 
effect of gravitational time dilation as studied here, should be included any 
model of nuclear matter within the neutron stars.

\emph{Acknowledgments:} SM would like to thank IISER Kolkata for supporting 
this work through a doctoral fellowship.

%
%

\end{document}